# Leakage Tests for Am-241 Solid Sources Used for Liquid Xenon Detector Monitoring

S. Bazzarri, G. Cicoli, P. De Felice[*], G. Rossi, F. Sedda

*ENEA C.R. Casaccia, C.P. 2400 I-00100 Rome - Italy.*

**Abstract.** Experimental tests are described to check solid $^{241}$Am radioactive sources used to monitor operation of liquid xenon detectors. In particular, radioactive leakage was tested in extreme temperature conditions following immersion in liquid nitrogen for different time periods. No radioactivity loss was detected from the sources considered. The paper describes the source characteristics, test methods and results obtained.

*Keywords:* Radioactive leakage test; check source; liquid xenon detectors.

## 1. Introduction

Alpha and beta solid radioactive sources are more and more used as check/monitor sources for liquid xenon detectors [1, 2]. The MEG experiment [3, 4] is the first that extensively used alpha sources distributed within the sensitive volume of a 100 L prototype liquid Xenon calorimeter. Larger calorimeters (800 L) are under construction. In this experiment, several point sources were mounted onto thin gold-plated tungsten wires. A number of these wires were fixed to the calorimeter walls and fully immersed in liquid Xe (165 K) for long period of time. A number of other different sources from two manufacturers [5, 6] were also used in preliminary work with prototype Xenon calorimeters.

Radioprotection requirements are then established to prevent significant radioactive contamination of the liquid scintillator. Furthermore, if such a contamination should occur, it would results in an increase of the intrinsic scintillation background with unwanted loss of sensitivity.

Stringent tests for leakage and contamination are essential feature for all radioactive source production [7]. The test method adopted depends on the design and intended application of the source and, also, on statutory requirements. For the specific intended use of Xe detector check sources, leakage tests normally required for solid sealed sources [7], cannot be exhaustive enough. The extremely low temperature conditions could introduce different mechanical and chemical-physical effects giving rise to a loss of radioactive material.

To meet these particular requirements, additional modified tests are needed, in particular to quantify the degree of radioactive leakage from solid radioactive sources following immersion in cold liquids, before their use in undergoing experiments. These test were requested by the Italian part of the MEG collaboration (INFN) to the National Institute for Ionising Radiation Metrology of ENEA (INMRI-ENEA) and are described in this paper.

---

[*] Corresponding author. E-mail address: *defelice@casaccia.enea.it*

## 2. Source characteristics

Tests were performed on three different types of sources. Sources a) and b) were made on small gold plated disks, 5 mm and 25 mm diameter respectively. Source c was made on 100 cm long thin wire. Main characteristics of these sources are reported in Table I.

*Table I – Characteristics of the tested sources.*

| Source | a | b | c |
|---|---|---|---|
| Manufacturer | Sorad Ltd. [5] | AEA Technology QSA GmbH [6] | Sorad Ltd. [5] |
| Manufacturer product code | APZ-1-J | AMR01032-LW 516 | - |
| Manufacturer drawing code | EL 819 | VZ-1366 | - |
| Intended use | Smoke detector | Alpha Wide Area Reference Source | Specific for MEG experiment |
| Nuclide | Am-241 | Am-241 | Am-241 |
| Nominal activity | 3 kBq | 3 kBq | 1 kBq |
| Leakage and contamination test performed by the manufacturer | [7], n. 5.1.4 | [7], n. 5.3.1 (wipe test) | - |
| Sealing technology | Covering foil | Incorporation in aluminium foil | Covering foil |
| Source geometry | Disk | Disk | Thin wire |
| Diameter of active surface (mm) | n.a. | 16 | - |
| Source dimensions (mm) | 5.1 (diameter) 0.2 (thickness) | 25 (diameter) 3 (thickness) | 1000 (length) 0.1 (diameter) |
| Source construction | • Silver backing (0.2 mm) • Gold interface (0.001 mm) • Active matrix (0.001 mm) • Gold cover layer (0.0016 mm) | • Aluminium holder • Anodised aluminium foil (0,3 mm thickness) with incorporated active layer (0.006 mm) | • Gold plated steel wire • Am-241 dots mounted by thermocompression • Gold cover layer |

The sources were preliminary tested for leakage and contamination by the manufacturer according to usual ISO requirements [7].

## 3. Test methods

The following additional tests were applied at INMRI-ENEA, on each source sequentially:

a) **Immersion in acetone:** the source is immersed in acetone bath at 20 °C for 6 hours, the removed activity is measured;
b) **Wiping (only for sources a and b):** the source is wiped with a paper filter, moistened with water, the removed activity is measured;
c) **Fast immersion in liquid nitrogen:** the source is repeatedly immersed 5 times in liquid nitrogen bath at 77 K for 3 min each, the removed activity is measured;
d) **Long immersion in liquid nitrogen:** the source is immersed in liquid nitrogen bath at 77 K for 1 hour, the removed activity is measured.

The tests were designed to check radioactive leakage under different conditions. Liquid nitrogen was used instead of xenon for immersion tests as it is characterised by a lower temperature (77 K) and easier availability.

Immersion tests were made in a polystyrene cubic vessel, 70 mm side, internally covered with a circular thin aluminium foil, 50 mm diameter. After complete evaporation of the liquid, the aluminium foil was removed and opened. This was then alpha-counted to check for any radioactivity leakage from the source.

Each aluminium foil was counted 5 times before and after each test. The three sources were also counted to detect any possible effect of the treatment on the particle emission rate. Background and blank measurement were performed between two foil counts. Standard sources, calibrated by the INMRI-ENEA, were used for calibration and for repeated counting system stability checks.

Alpha counting was performed on a Thermo Eberline ESM FHT 650 K1 low background counting system equipped with an active anticoincidence shield (Table II). Constant counting times of 6000 s (sources a and b) and 3000 s (source c) were imposed for background, blank and source counting. Sources c were measured in a building other than that for sources a and b, and characterised by lower and more stable environmental background.

*Table II – Characteristics of the counting system.*

| Model | Thermo Eberline ESM FHT 650 K1 |
|---|---|
| Type | Low background alpha-beta counter |
| Measurement detector | Flow-type proportional counter tube |
| Guard detector | Flow-type proportional counter tube |
| Gas | Argon-Methane (P10) |
| Window material | Aluminized Mylar |
| Window diameter | 60 mm |
| Window thickness | 0.3 mg/cm$^2$ |
| Counting efficiency | 28% (Am-241 $\alpha$ particles) |

## 4. Results

Blank and background (BB) count rates were not distinguishable as well as those for aluminium foils and paper filters (AP). For simplicity they were then grouped in the following description of the test results. Both BB and AP readings were not Poisson distributed, being other sources of variation present other than pure statistical fluctuations.

*Table III – Test conditions and results.*

| Source | a and b | c |
|---|---|---|
| Number of BB readings | 64 | 13 |
| Counting time (s) | 6000 | 3000 |
| Mean BB count rate (s$^{-1}$) | 0.015 | 0.0042 |
| Experimental Standard Deviation of the BB count rate (s$^{-1}$) | 0.0072 | 0.0013 |
| Mean AP count rate (s$^{-1}$) | 0.0122 | 0.0040 |
| Experimental standard deviation of the AP count rate (s$^{-1}$) | 0.0062 | 0.0016 |
| Decision threshold for the net individual count-rate (s$^{-1}$) | 0.012 | 0.0022 |
| Decision threshold for the net mean count-rate (s$^{-1}$) | 0.0055 | 0.0011 |
| Detection limit for the net mean count-rate (s$^{-1}$) | 0.011 | 0.0023 |
| Minimum detectable activity (Bq) | 0.039 | 0.008 |
| Measured leakage factor | < 6.6 10$^{-6}$ | < 4.0 10$^{-6}$ |
| Minimum detectable leakage factor | 1.3 10$^{-5}$ | 8.1 10$^{-6}$ |

The mean and experimental standard deviation of BB and AP count rates were then obtained from repeated readings. The mean BB count rate was subtracted from any individual AP count rate value. Decision thresholds and detection limits were calculated for the individual and mean count rate values, according to relevant ISO recommendations [8], taking into account the different degrees of freedom for BB and AP readings and assuming a level of confidence of 95% ($\alpha=\beta=0.05$). The results obtained for sources a, b and c are reported in Table III.

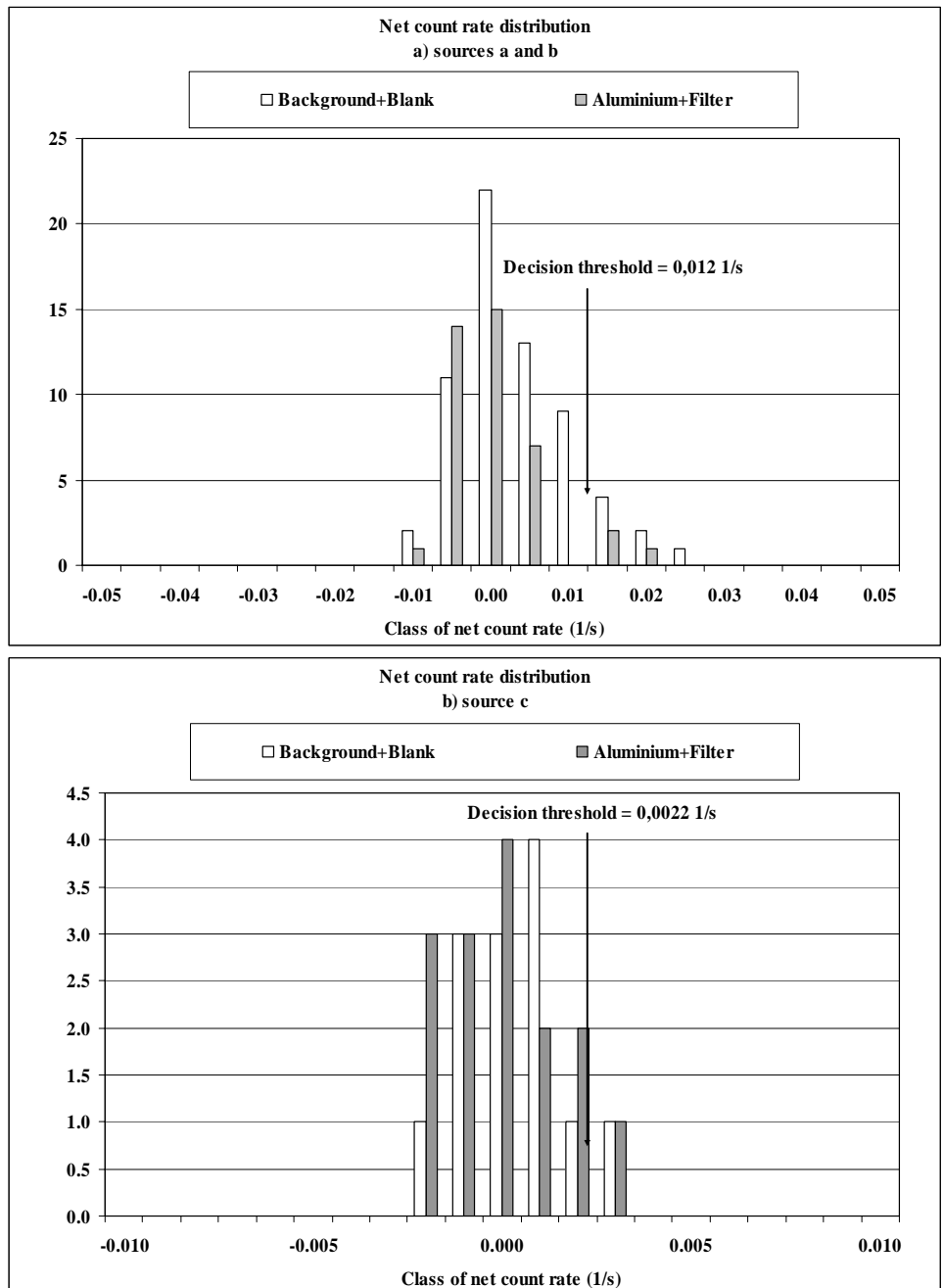

*Figure 1 – Combined distributions of individual net count-rate values for BB and for AP, for sources a, b (graph a) and c (graph b). No significant difference between the BB and AP distributions can be observed. Statistical analysis confirms this intuitive conclusion. Almost the same fraction of BB and AP results overcomes the decision thresholds.*

The BB and for AP net count-rate distributions are reported in Fig. 1 and 2. No significant difference can be found between the two distributions. No one mean AP count rate value resulted higher than the respective decision threshold.

Repeated measurements of disk sources did not reveal any change in the counting efficiency confirming the sources stability.

This results were then considered satisfactory for the intended use of the sources.